\documentclass[11pt]{article}

\usepackage{epsfig}
\usepackage{fancyhdr}
\usepackage{color}
\usepackage{verbatim}
\usepackage{lineno}
\usepackage{cite}
\usepackage{caption}
\usepackage{url}
\usepackage{graphicx}
\usepackage{comment}
\usepackage[hmargin=3cm,vmargin=3.0cm]{geometry}

\begin{document}
\title{CRT: A Numerical Tool for Propagating Ultra-High Energy Cosmic Rays Through Galactic Magnetic Field Models}
\author{M.S. Sutherland\footnote{msutherl@mps.ohio-state.edu}, B.M. Baughman\footnote{bbaugh@mps.ohio-state.edu}, J.J. Beatty\footnote{beatty@mps.ohio-state.edu}\\
The Ohio State University\\
Columbus, Ohio, USA 43210
}
\date{2010}
\maketitle

\abstract{Deflection of ultra high energy cosmic rays (UHECRs) by the Galactic magnetic field (GMF) may be sufficiently strong to hinder identification of the UHECR source distribution.
A common method for determining the effect of GMF models on source identification efforts is backtracking cosmic rays.
We present the public numerical tool \textit{CRT} for propagating charged particles through Galactic magnetic field models by numerically integrating the relativistic equation of motion.
It is capable of both forward- and back-tracking particles with varying compositions through pre-defined and custom user-created magnetic fields.
These particles are injected from various types of sources specified and distributed according to the user.
Here, we present a description of some source and magnetic field model implementations, as well as validation of the integration routines.}

\section{Introduction}
\label{sec:introduction}
Ultra-high energy cosmic rays are thought to be charged protons or nuclei, due to constraints on the photon flux at the highest energies \cite{2009APh....31..399T}.
These particles will therefore be deflected by galactic and extragalactic magnetic fields during propagation from their sources.
Any type of UHECR source identification must account for this deflection.
Studies including Stanev \cite{1997ApJ...479..290S}, Medina Tanco et al. \cite{1998ApJ...492..200M}, Harari et al., \cite{1999JHEP...08..022H}, Yoshiguchi et al. \cite{2003ApJ...596.1044Y}, Prouza and {\v S}m{\'{\i}}da \cite{2003A&A...410....1P}, Takami et al. \cite{2006ApJ...639..803T}, and Kachelrie{\ss} et al. \cite{2007APh....26..378K} have used the method of backtracking cosmic rays through various GMF models as a source discrimination tool.
However, these studies investigated particular configurations of the GMF with fixed quantities for parameters such as the overall normalization and scale heights.
Additionally, the propagation codes used in each study were individually developed making a direct comparison of the propagation accuracy between the codes difficult.

Two commonly cited public propagation codes for cosmic rays are GALPROP\footnote{http://galprop.stanford.edu/web$\_$galprop/galprop$\_$home.html} \cite{1998ApJ...509..212S} and CRPropa\footnote{http://apcauger.in2p3.fr/CRPropa/} \cite{2007APh....28..463A}.
However, the focus and design of these codes differ from \textit{CRT}, which focuses on minimizing computation time for tracking UHECRs of arbitrary composition.
These codes are designed to investigate the effects of interaction and energy loss during cosmic ray propagation.
While GALPROP contains a detailed exponential model of the GMF, particles are propagated by solving the diffusion equation which is more suited for low energy (GeV-TeV) cosmic ray studies in the GMF.
CRPropa numerically integrates the equations of motion but implements only extragalactic turbulent magnetic fields.
In particular, the CRPropa code is not optimized for convoluted GMF models and only propagates cosmic rays with $Z=1$.

Presented here is a public tool for propagating UHECR events through various models of the GMF.
This program implements components of field models used in previous studies and many possible source distributions.
The user is capable is specifying many simulation parameters such as tracking method (forward or back), magnetic field model and source configurations, and integration routine settings.
This is accomplished through a simply-designed interface.
Output includes useful information about the injection and observation site coordinates and deflection magnitude.
Header information in the output file provides a record of the simulation settings.
The code is actively maintained with revision control using subversion\footnote{http://subversion.tigris.org}.
A website\footnote{http://crt.mps.ohio-state.edu/} containing public releases of the code is accessible under credentials\footnote{username: \textit{public}, password: \textit{tracker}}.
Complete simulation information beyond that described here can be found in README text files accompanying each version.

\section{Overview}
\label{sec:overview}
The numerical tool \textit{CRT} is written in C++ and has only a single dependency on the GSL random number generation libraries \cite{gslwebsite}.
\textit{CRT} is capable of forward- and back-tracking cosmic rays from their injection sources to a detector or magnetic field boundary.
Command line options set the global parameters for a specified program run, such as backtracking status and detector size.
The desired injection sources and magnetic field models, as well as their parameter values, are specified by the user in an external configuration file processed upon program execution.

UHECR events are injected at user-specified sources and propagated through a user-specified magnetic field by integrating the relativistic Lorentz equation.
If the backtracking mode is selected, the source is located at the Galactic position of the Sun within the GMF model.
\textit{CRT} implements $5^{th}$ order adaptive Runge-Kutta integration routines for determining the cosmic ray trajectories.
The term ``adaptive'' arises from checks performed prior to completing an integration step.
The step is recalculated with a smaller stepsize if the induced errors are larger than the user-specified tolerance.
These routines are more fully described in Numerical Recipes \cite{1403886}.
Trajectories can be recorded step-by-step or at specific locations\footnote{Changing the recording method simply requires recompiling with a different preprocessor command. This is done to reduce runtimes.}, such as the boundary of the detector.
Cosmic rays are propagated until they reach the boundary of the field model, integration time limit, or detector.

Energy losses are not considered during propagation.
The GZK process \cite{1966PhRvL..16..748G, 1966JETPL...4...78Z} involves cosmic ray interactions with the cosmic microwave background (CMB) resulting in photo-pion production for protons or photodisintegration of heavy nuclei and is the dominant energy loss on UHECRs due to the CMB constituting the dominant particle species in extragalactic space.
The same process can occur within the Galaxy due to interactions with the ambient starlight and infrared radiation from stars and warm dust.
However in the Galactic center, the infrared photon density for energies above threshold energy is estimated to be of order 1 cm$^{-3}$ or less (see, e.g., \cite{2005ICRC....4...77P}) and decreases with galactocentric distance.
The CMB remains the dominant photon field $\left(n_{cmb}=411\ \mbox{cm}^{-3}\right)$ even in galactic interstellar space, but UHECR interactions are infrequent due to the reduced pathlength.
UHECRs may travel of order tens of kpc through the Galaxy as opposed to Mpc in intergalactic space.
Above threshold energy, the interaction cross-section $\sigma_{p\gamma}$ is roughly constant at $10^{-1}$ mb.
The probability for a single interaction is $P = L n_{cmb} \sigma_{p\gamma}$, so for a typical pathlength through the Galactic center $L\approx10$ kpc, $P\approx10^{-3}$.
Effects from energy loss by synchrotron emission due to the strong magnetic fields observed in the Galactic center region ($B\approx$ tens of mG) (see, e.g., \cite{2001SSRv...99..243B, 2010Natur.463...65C}) are also negligible at EeV energies and higher.
Energy loss due to inverse Compton scattering is unimportant at ultra-high energies.

\begin{figure}[t]
\begin{center}
\includegraphics[width=0.45\textwidth]{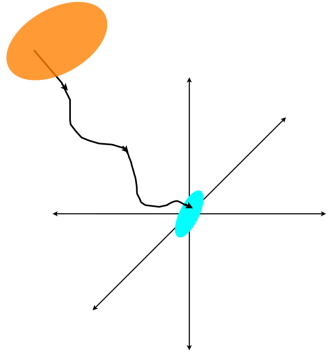}
\caption{\label{fig:discs}Detection (blue) and injection (orange if forwardtracking; blue if backtracking) disc implementation in \textit{CRT}.
This image depicts a scenario where a cosmic ray is forwardtracked and detected from an extragalactic source.
The black line traces the particle trajectory from the injection disc to the detection disc.}
\end{center}
\end{figure}

\textit{CRT} implements a single detector scheme.
It is constructed as a disc located at the Sun's position in the Galactic plane and oriented such that the surface normal always lies parallel to the particle trajectory.
The detector is shown as a blue disc at the center of the coordinate system in Figure \ref{fig:discs}.

\begin{figure}[ht]
\begin{center}
\includegraphics[width=0.45\textwidth]{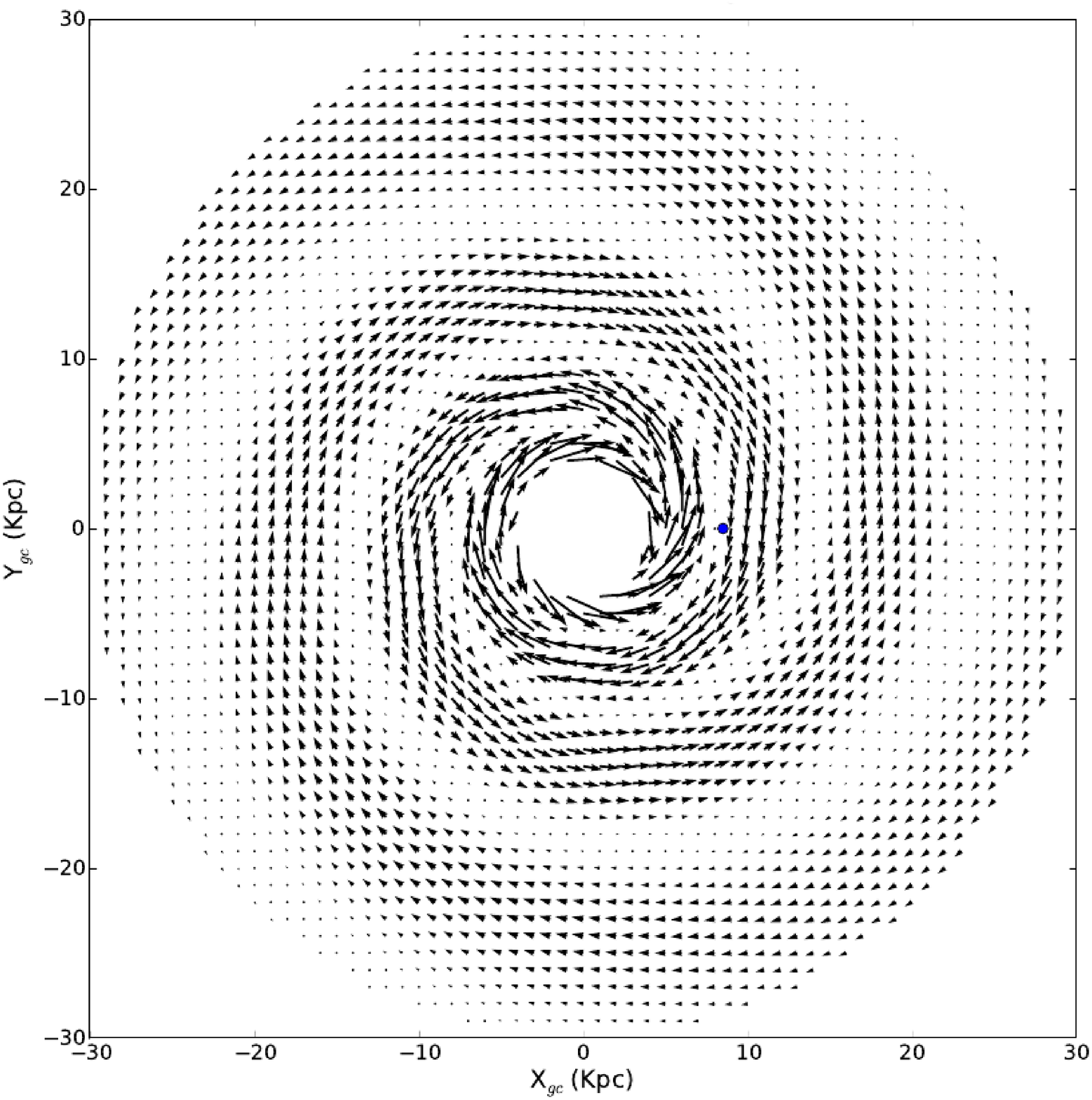}
\caption{\label{fig:bss-field-structure}HMR BSS \cite{1999JHEP...08..022H} field structure in the Galactic plane (z=0) for galactocentric distances $r_{gc}\leq30$ kpc.
Field vectors for $r_{gc}=4$ kpc are masked and not shown.
The blue dot at $(x_{gc},y_{gc})=(8.5\ \mbox{kpc},\ 0)$ represents the solar position.}
\end{center}
\end{figure}

\textit{CRT} also includes additional executables specifically designed for testing the integration routines and parameterizing the magnetic field models.
Use of the integration accuracy test program is demonstrated in Section \ref{sec:valid}.
For a user-specified GMF configuration, the executable \verb-getfield- outputs information about the magnetic field useful for producing three dimensional grid plots of the field strength and direction.
An example plot is shown in Figure \ref{fig:bss-field-structure}.

\section{Magnetic Field Models}
\label{sec:fields}
The public releases of \textit{CRT} incorporate descriptions of many magnetic field models, but is modularly designed so that the user may easily write and include additional field models, a central feature in its design.
It is possible to use any number and combination of default and user-written models simultaneously during the execution of the program; the user specifies the desired configuration in the external configuration file.

Field models included in the public releases are: Logarithmic Spiral, Dipole, Toroidal, Ring, and a simple random field.
The user is capable of specifying every parameter value of these field models.
These individual field models can be combined to create more complex field models, such as that studied by Prouza and {\v S}m{\'{\i}}da \cite{2003A&A...410....1P} for example, and also combined with any user-created field models.
These default models are fully described in documentation accompanying each release version.
The use of these magnetic field models is entirely optional by the user.
Only models specified by the user in the external configuration will be used during program execution.

Poloidal and Toroidal fields are thought to exist as components of the Galactic halo field away from the Galactic plane and so are included.
Logarithmic Spiral field models have seen extensive use in previous cosmic ray and radio starlight studies (see, e.g., \cite{1997ApJ...479..290S, 1999JHEP...08..022H, 1998ApJ...492..200M, 2003A&A...410....1P, 2008A&A...477..573S, 2009JCAP...07..021J}).
The Logarithmic Spiral field model incorporates different symmetry parameterizations about the Galactic rotation axis and plane, respectively, and radial dependencies from two commonly studied field model : Stanev \cite{1997ApJ...479..290S} and Harari-Mollerach-Roulet (HMR) \cite{1999JHEP...08..022H}.
In total, \textit{CRT} includes the descriptions of 8 unique types of Logarithmic Spiral models available for use.

The Galactic magnetic field also appears to have a significant random component (see, e.g., \cite{2001SSRv...99..243B, 2008A&A...477..573S}), which can be approximated with a simple random field model.
The simple random field in \textit{CRT} is constructed by populating individual spherical magnetic cells each with a single direction and magnitude chosen from gaussian distributions.
The gaussian distributions are described by parameters specified by the user.
Cell boundaries are respected during individual cosmic ray propagation, ensuring a proper magnetic field calculation at each position.

It is important to reiterate that the user has complete freedom in choosing which magnetic field models to include during program execution.
The user could select every field model listed above, or two, or even none and instead choosing to recompile \textit{CRT} with their own source code.
The creation of any user-written field model simply requires constructing and specifying the field model in a fashion similar to the included models.
The construction can be as simple as amending the radial dependence of the field strength equation of a particular default spiral model or a user implementing a highly complex field model in new source code.

\section{Sources and Spectra}
\label{sec:sources}
\textit{CRT} includes many possible source descriptions allowing for a variety of source distributions.
As with the magnetic field models, \textit{CRT} is capable of incorporating any combination of included and user-written sources and energy spectra during the program execution.
The user is capable of specifying all parameter values of every source type.
All sources inject cosmic rays with energies sampled from a power law distribution,
\begin{equation}
\frac{dN}{dE_{inj}} \propto E^{-\alpha}_{inj}.
\end{equation}
\noindent Some source types are capable of sampling energies from a broken power law distribution described by two spectral indices and a break energy.
Cosmic rays injected from any source type are capable of both forward- and back-tracking.

One of the included source types, Isotropic sources generate events with initial velocity directions sampled isotropically across the sky.
Once a direction has been selected, the injection site is determined within a disc of user-specified size lying within the tangent plane to that direction.
In the case of forward-tracked cosmic rays, events are injected just beyond the magnetic field model boundary from a random location on the disc in order to mimic an extragalactic flux.

The user is capable of specifying the sizes of the injection disc for each source in the external configuration file and the detection disc as a runtime command line parameter.
In a forward tracking scenario, injection disc radii larger than roughly kpc effectively mimic an impinging extragalactic cosmic ray flux on the Galaxy.
Detector discs with radius of order 0.1 kpc efficiently detect forwardtracked events. 
The Sun is estimated to be roughly 20 pc above the Galactic plane, as inferred from stellar surveys (see, e.g., \cite{1995AJ....110.2183H, 2007MNRAS.378..768J, 2009MNRAS.398..263M}).
This detector radius value accounts for the error on the Sun's galactic position and gives a good compromise between accumulating detected events and properly simulating the local GMF.
Figure \ref{fig:discs} illustrates the orientation of the discs during runtime.
For a simulation incorporating forwardtracking, the orange disc represents the injection disc of an extragalactic source.
A particle trajectory is shown originating from a randomly selected position on the injection disc, propagating through a magnetic field, and impacting the detector disc.
At each integration step, the detector disc is oriented such that its normal is parallel to the particle trajectory.

The Isotropic source type is one of many included source types in the public release.
Additional source types are capable of injecting particles from any location or adjusting the injection direction to account for estimated deflection by a simple extragalactic turbulent magnetic field model\footnote{\textit{CRT} does not propagate the particles through extragalactic space. For small scale fields, the deflection manifests as a gaussian smearing of the cosmic ray directions.}.
There are also source types that incorporate sky coverage of earth-based detectors\footnote{Available only for backtracking.}.
This coverage depends on the declination and limiting zenith angle of the detector, which can be easily changed within the source code.
These sources are identical to other source types but their source code incorporates the additional coverage factor in selecting event directions.

All source types are more fully described on the website and in documentation accompanying the source code.

\section{Validation}
\label{sec:valid}
\begin{figure}[p]
\centerline{\includegraphics[width=0.45\textwidth]{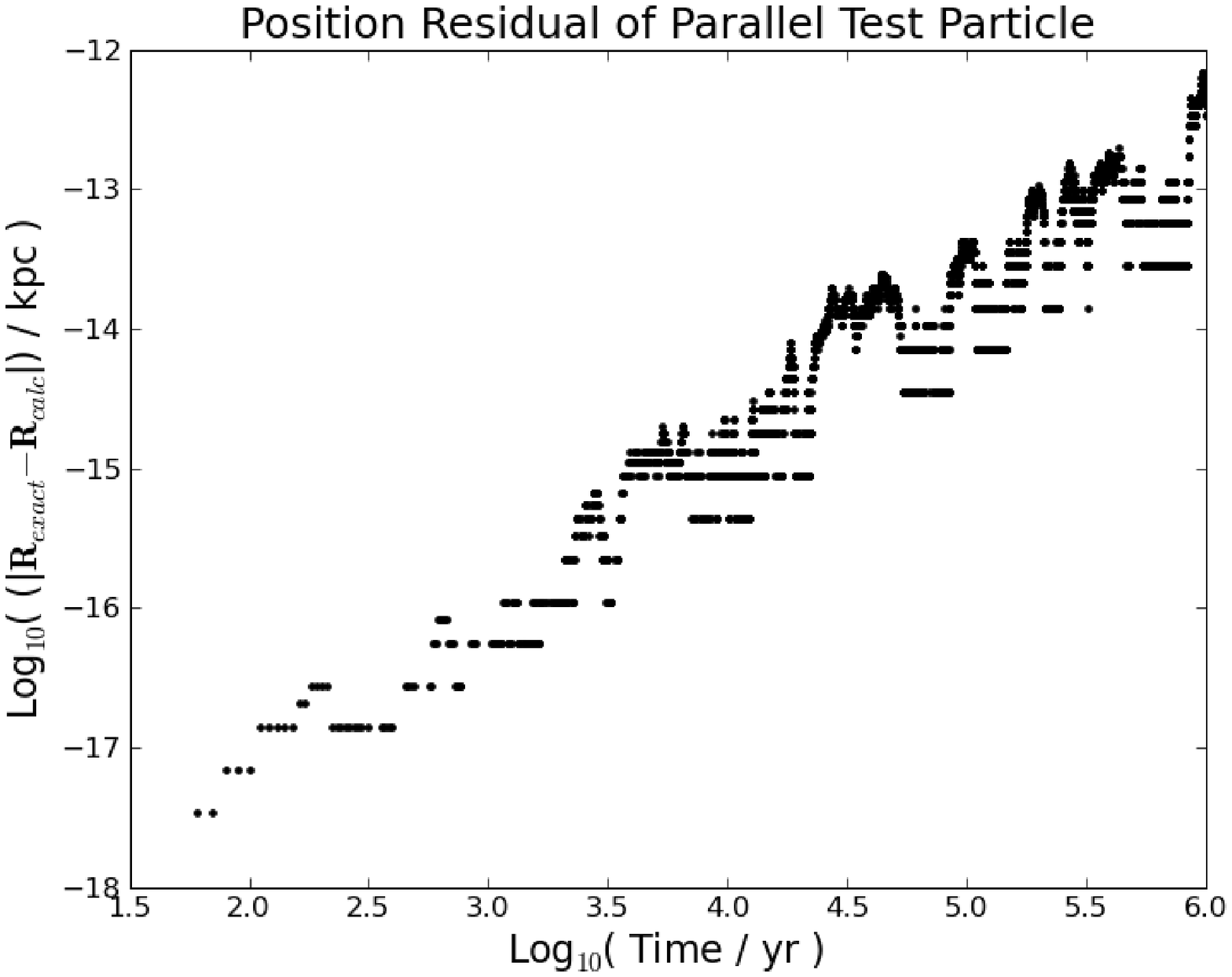}
            \hfil
            \includegraphics[width=0.45\textwidth]{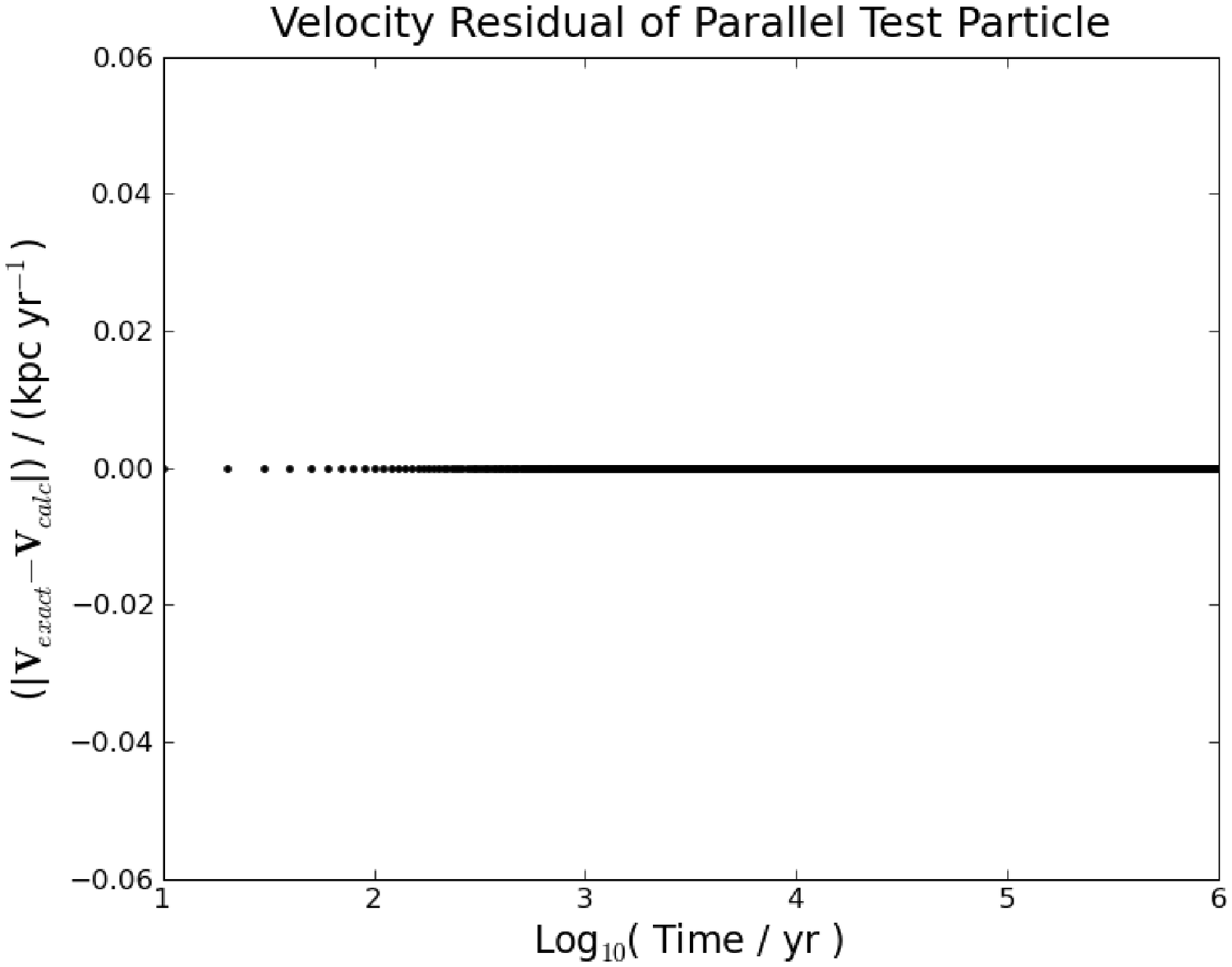}
           }
\caption{\label{fig:parallel}Residuals for a particle initially traveling parallel to the magnetic field.
The left plot shows the position residual and the right plot shows the velocity residual.}
\end{figure}

\begin{figure}[p]
\centerline{\includegraphics[width=0.45\textwidth]{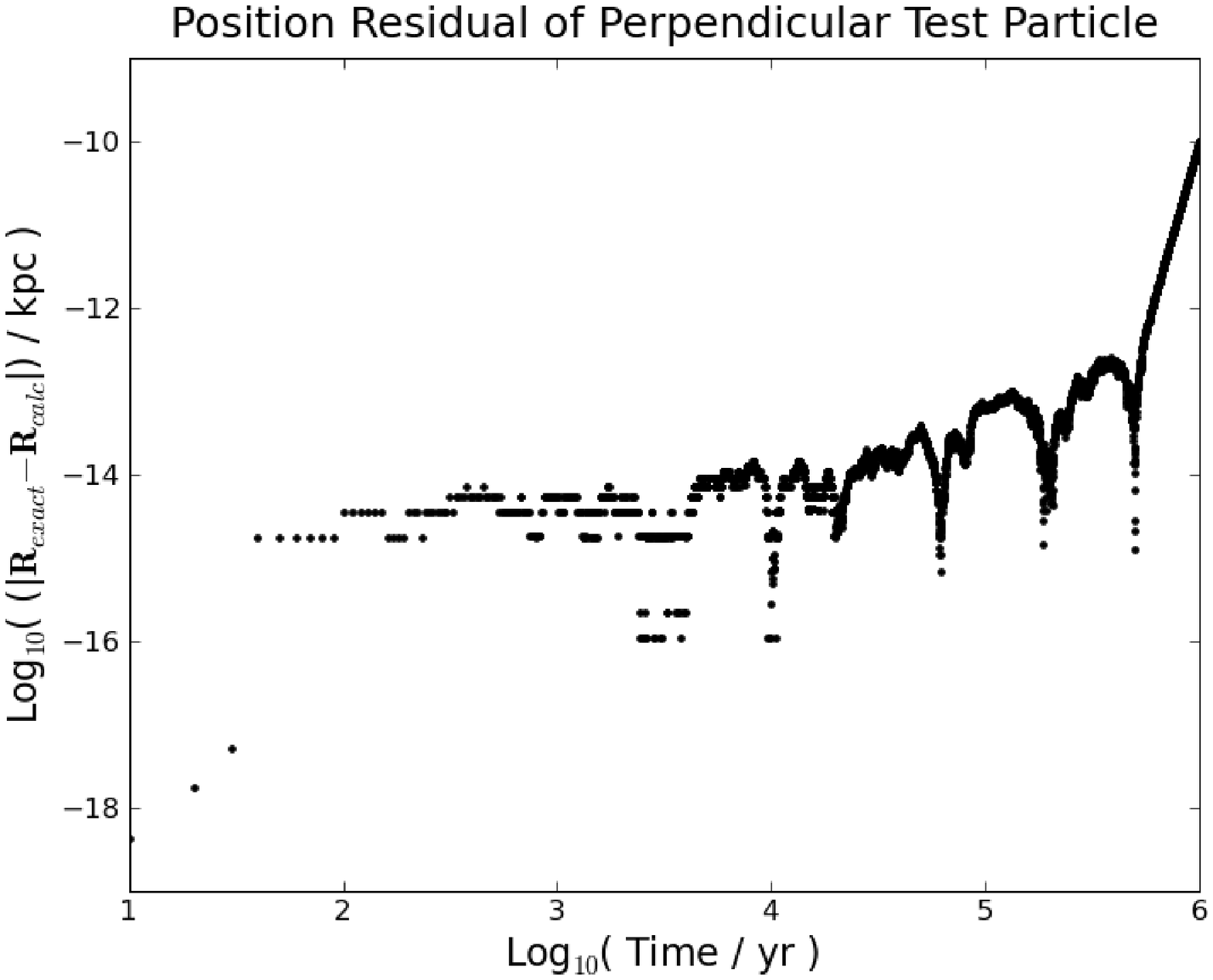}
            \hfil
            \includegraphics[width=0.45\textwidth]{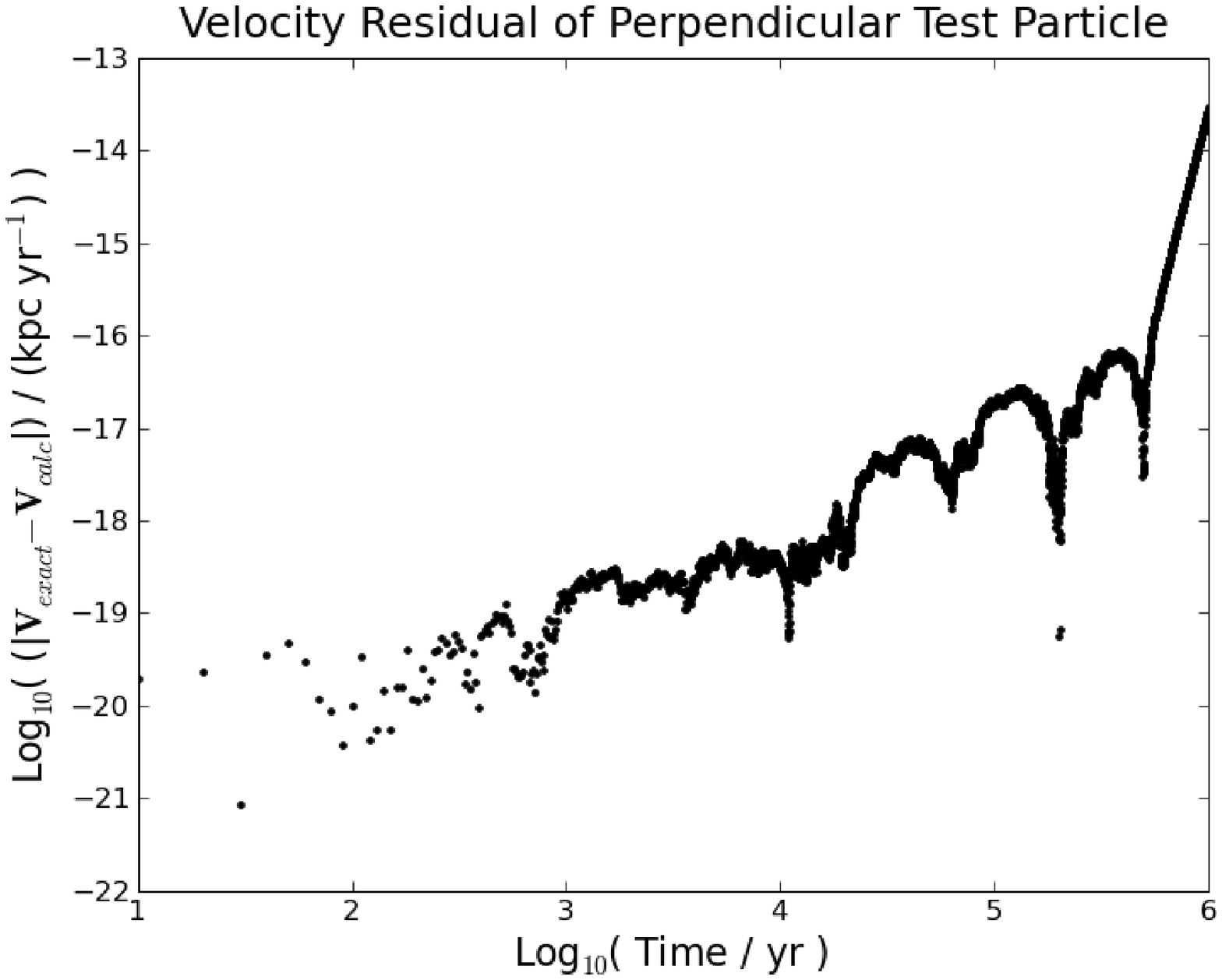}
           }
\caption{\label{fig:perpendicular}Residuals for a particle initially traveling perpendicular to the magnetic field.
The left plot shows the position residual and the right plot shows the velocity residual.}
\end{figure}

The accuracy of the integration routines has been tested by tracking charged particles in a uniform field.
Particles traveling initially perpendicular to the field will execute simple circular motion whereas particles traveling initially parallel will propagate undeflected.
The integration accuracy can be estimated by determining the difference between the exact and calculated positions and velocities as functions of time.
The residuals are $dR=|\vec{R}_{exact}-\vec{R}_{calc}|$ and $dV=|\vec{V}_{exact}-\vec{V}_{calc}|$.
Figures \ref{fig:parallel} and \ref{fig:perpendicular} show the residuals as functions of time for 1 EeV protons injected into a uniform $1$ $\mu$G magnetic field.
The initial velocity vectors are parallel and perpendicular to the magnetic field, respectively.

In the parallel injection case (Figure \ref{fig:parallel}), the particle propagates undeflected and the velocity residual always zero.
The position residual at the beginning of propagation is dominated by numerical accuracy.
In this simulation, the particle travels parallel to a coordinate axis, so that at large times, the exact and calculated coordinate values are each large values.
The calculation of the residual is the difference of two large values, leading to rounding uncertainty at large times.

The perpendicular case (Figure \ref{fig:perpendicular}) depicts a different scenario.
The velocity residual shows the accumulated effect of numerical accuracy over time.
The residual decreases at particular times, but generally shows a trend toward increasing values.
The position residual is directly influenced by the velocity residual since the integration routine uses the calculated velocity values to determine the position.
The behavior of the position residual plot closely mimics that of the velocity residual with negligible time delay.
The residual value is initially near numerical accuracy, then slowly begins increasing around the same time as the velocity residual.

The position residuals in both injection cases are well below $10^{-10}$ for an integration time of $10^{6}$ yr, during which the total pathlength of the particles is roughly 300 kpc.
The diameter of the Milky Way as resolved by its distribution of stars and gas is about 20-30 kpc and the magnetic field is thought to be roughly the same size.
The Larmor radius for a UHECR in a uniform field is,
\begin{equation}
R = 1.08 \left(\frac{1}{Z}\right) \left(\frac{E}{\mbox{EeV}}\right) \left(\frac{\mu\mbox{G}}{B}\right)\ \mbox{kpc},
\end{equation}
\noindent where $Z$ and $E$ are the cosmic ray charge and energy in EeV, and $B$ the field strength in $\mu$G.
At $E\approx 10$ EeV the Larmor radius becomes comparable to the Galactic diameter and only high charge values or magnetic fields with considerable field strengths can contain the UHECRs.
In realistic GMF models, the field vector varies significantly in strength and direction over length scales of $10^{2-3}$ pc such that deflection induced from one region of magnetic field may be destructively interfered by the next region.
These field models will influence the cosmic ray trajectory in a manner somewhat similar to random fields, so that the trajectory still maintains quasi-rectilinear behavior even at relatively low energy ($E\approx$ few EeV).

\section{Example Simulations}
\label{sec:sims}
Here we illustrate some useful applications of the output of \textit{CRT}.
Upon completion of tracking a particle, information is recorded including the initial and final conditions of the particle, the deflection magnitude, source properties, and detection status.
The header information contains complete information about the simulation settings as well as the program call at the command line.

In the first example, $10^{4}$ protons are backtracked from an isotropic observed distribution with a differential energy spectrum $dN/dE\approx E^{-3}$ with minimum and maximum energies $E_{min}=50$ EeV and $E_{max}=200$ EeV.
The Galactic magnetic field is represented by the HMR BSS$\_$A configuration shown in Figure \ref{fig:bss-field-structure}.
The field experiences exponential attenuation away from the Galactic plane with two scale heights: 0.5 kpc and 3.5 kpc.

\begin{figure}[p]
\begin{center}
\includegraphics[width=0.65\textwidth]{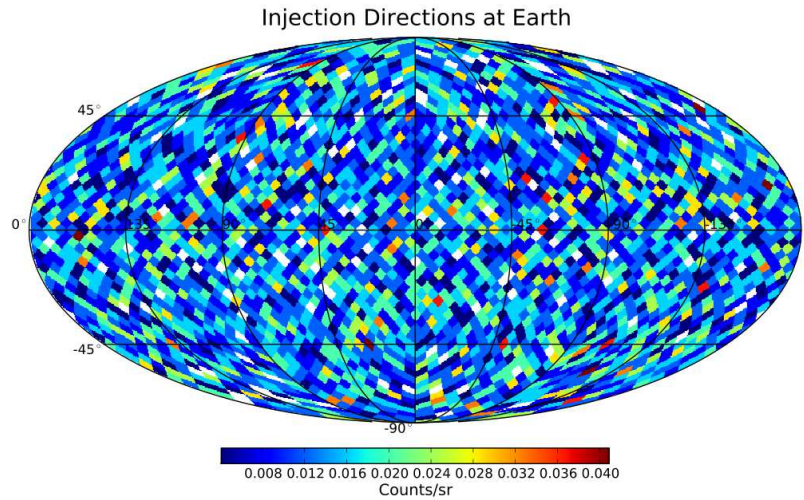}
\caption{\label{fig:scenario1-observed-skymap}Injection directions for events in the backtracked example.
The map is shown in Galactic coordinates using the HEALPIX package \cite{2005ApJ...622..759G}}
\end{center}
\end{figure}

\begin{figure}[p]
\begin{center}
\includegraphics[width=0.65\textwidth]{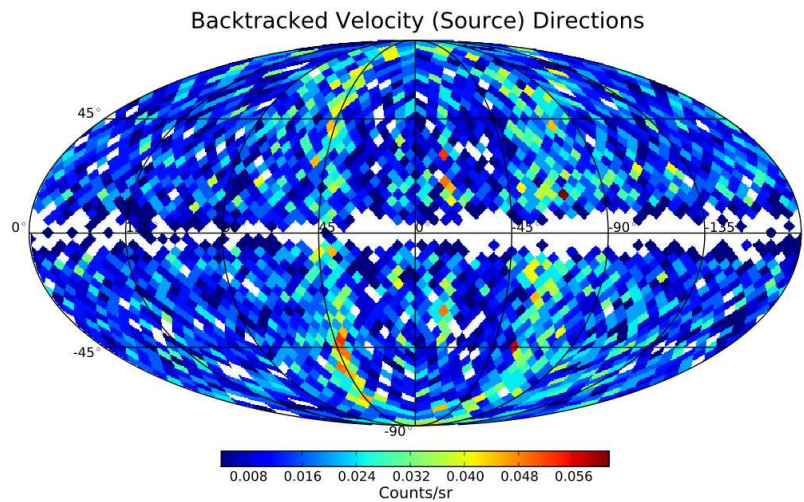}
\caption{\label{fig:scenario1-source-skymap}Reconstructed source directions inferred from the velocity direction after backtracking.
The map is shown in Galactic coordinates using the HEALPIX package \cite{2005ApJ...622..759G}}
\end{center}
\end{figure}

\begin{figure}[p]
\begin{center}
\includegraphics[width=0.55\textwidth]{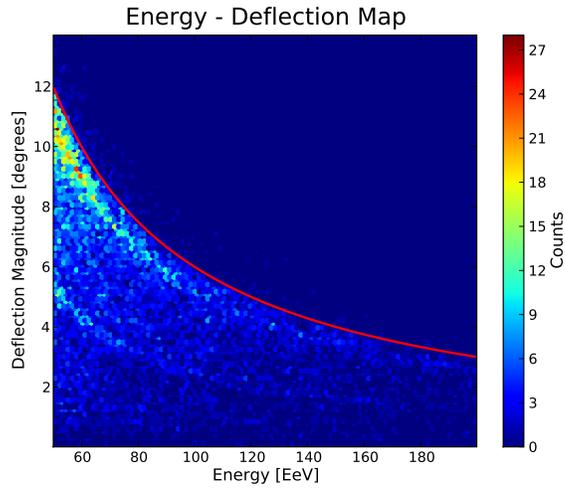}
\caption{\label{fig:scenario1-energy-deflection}Heat diagram of the energy and deflection magnitude for each particle in the backtracked example.
The red line indicates $E^{-1}$ and is meant to guide the eye.}
\end{center}
\end{figure}

Figure \ref{fig:scenario1-observed-skymap} shows a Hammer-Aitoff projected skymap in Galactic coordinates of the density of injection directions.
These directions are selected from an isotropic sky distribution.
Figure \ref{fig:scenario1-source-skymap} indicates the inferred source direction of the particles based on the direction of the velocity vector upon crossing the magnetic field model boundary.
Structure is apparent at specific Galactic longitudes, particularly $\ell \approx 45^{\circ}, \pm10^{\circ}, -50^{\circ}$, due to the semi-regular spiral structure towards the central region of the field model.
The nature of the field model induces the low energy particles to spiral along the magnetic field lines before exiting.
The magnetic field model also deflects nearly all particles out of the Galactic plane due to increased pathlength across the plane through regions of large field strength.
Figure \ref{fig:scenario1-energy-deflection} shows the deflection magnitude as a function of energy for all particles.
The red line serves to guide the eye along a curve following $E^{-1}$ behavior, such as that expected by energy-dependent deflection in a uniform field.
It is apparent at low energy that some events experience less deflection than would be expected, as these particles are lensed along magnetic field lines into the nearest structure seen in Figure \ref{fig:scenario1-source-skymap}.

\begin{figure}[p]
\begin{center}
\includegraphics[width=0.75\textwidth]{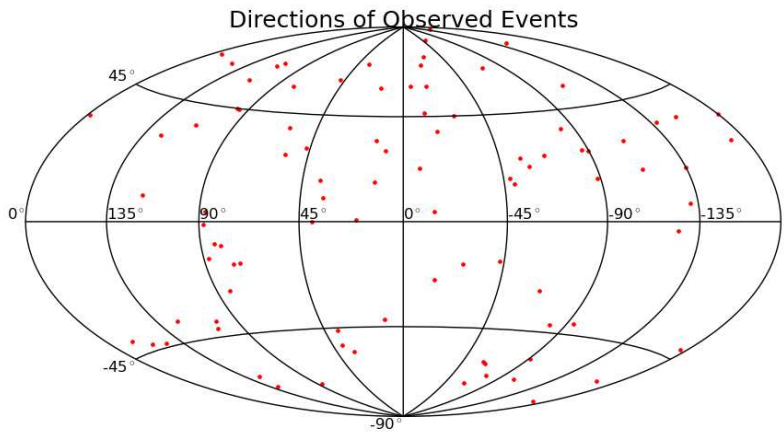}
\caption{\label{fig:scenario2-observed-skymap}Observed event directions in the forward-tracked example.
Only the 93 detected events are shown.}
\end{center}
\end{figure}

\begin{figure}[p]
\begin{center}
\includegraphics[width=0.45\textwidth]{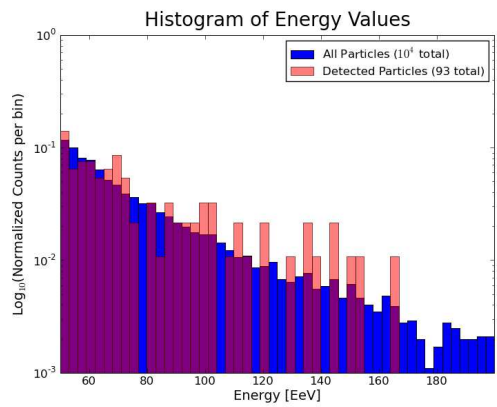}
\caption{\label{fig:scenario2-energy-histogram}Energy histogram for all simulated (blue) and detected (red) particles in the forward-tracked example.}
\end{center}
\end{figure}

In a second example, $10^{4}$ protons are forward-tracked from an extragalactic isotropic source distribution through the same magnetic field model as above.
The injection site for each particle is randomly selected within an injection disc of radius 1 kpc.
The detector radius is set to 0.1 kpc.
Figure \ref{fig:scenario2-observed-skymap} shows the sky distribution of observed events.
Figure \ref{fig:scenario2-energy-histogram} shows normalized count histograms of the energies for detected and non-detected events.

\section{Conclusion}
\label{sec:conclusion}
We have presented a publicly available numerical tool for tracking cosmic ray events from various source distributions through Galactic magnetic field models.
\textit{CRT} incorporates components of basic models of the Galactic magnetic field and many possible source distributions from which to create and propagate cosmic ray events.
The modularity of \textit{CRT} allows for users to easily implement modules for new magnetic field models and sources.
The integration accuracy of the program has been tested and deemed acceptable for most simulation configurations.


\bibliographystyle{JHEP-2}
\bibliography{document}

\end{document}